\documentclass[%
 reprint,
 amsmath,amssymb,
 aps, superscriptaddress
]{revtex4-2}

\usepackage{graphicx}
\usepackage{dcolumn}
\usepackage{bm}
\usepackage{hyperref}
\usepackage[caption=false]{subfig}

\usepackage[dvipsnames]{xcolor}
\usepackage{aas_macros}

\usepackage{orcidlink}

\usepackage[normalem]{ulem}
\definecolor{granadagreen}{HTML}{078931}

\begin{document}

\preprint{APS/123-QED}
\title{Enhancing Gravitational Wave Parameter Estimation with Non-Linear Memory: Breaking the Distance-Inclination Degeneracy}
\author{Yumeng Xu\,\orcidlink{0000-0001-8697-3505}}
\affiliation{%
 Physik-Institut, Universit\"{a}t Z\"{u}rich, Winterthurerstrasse 190, 8057 Z\"{u}rich, Switzerland
}

\author{Maria Rossell\'o-Sastre\, \orcidlink{0000-0002-3341-3480}}
\affiliation{
Departament de F\'{i}sica, Universitat de les Illes Balears, IAC3 – IEEC, Crta. Valldemossa km 7.5, E-07122 Palma, Spain}

\author{Shubhanshu Tiwari\,\orcidlink{0000-0003-1611-6625}}
\affiliation{%
 Physik-Institut, Universit\"{a}t Z\"{u}rich, Winterthurerstrasse 190, 8057 Z\"{u}rich, Switzerland
}
\author{Michael Ebersold\,\orcidlink{0000-0003-4631-1771}} 
\affiliation{%
 Laboratoire d'Annecy de Physique des Particules, CNRS, 9 Chemin de Bellevue, 74941 Annecy, France
}

\author{Eleanor Z Hamilton\,\orcidlink{0000-0003-0098-9114}}
\affiliation{
Departament de F\'{i}sica, Universitat de les Illes Balears, IAC3 – IEEC, Crta. Valldemossa km 7.5, E-07122 Palma, Spain}
\affiliation{%
 Physik-Institut, Universit\"{a}t Z\"{u}rich, Winterthurerstrasse 190, 8057 Z\"{u}rich, Switzerland
}

\author{Cecilio Garc\'ia-Quir\'os\,\orcidlink{0000-0002-8059-2477}} 
\affiliation{%
 Physik-Institut, Universit\"{a}t Z\"{u}rich, Winterthurerstrasse 190, 8057 Z\"{u}rich, Switzerland
}

\author{H\'ector Estell\'es\,\orcidlink{0000-0001-6143-5532}} 
\affiliation{%
 Max Planck Institute for Gravitational Physics (Albert Einstein Institute), Am M\"{u}hlenberg 1, Potsdam 14476, Germany
}

\author{Sascha Husa\,\orcidlink{0000-0002-0445-1971}}
\affiliation{
Institut de Ci\`encies de l'Espai (ICE, CSIC), Campus UAB, Carrer de Can Magrans s/n, 08193 Cerdanyola del Vall\`es, Spain}

\affiliation{
Departament de F\'{i}sica, Universitat de les Illes Balears, IAC3 – IEEC, Crta. Valldemossa km 7.5, E-07122 Palma, Spain}

\date{\today}

\begin{abstract}

In this study, we investigate the role of the non-linear memory effect in gravitational wave (GW) parameter estimation, particularly we explore its capability to break the degeneracy between luminosity distance and inclination angle in binary coalescence events. Motivated by the rapid growth in GW detections and the increasing sensitivity of GW observatories enhancing the precision of cosmological and astrophysical measurements is crucial.
We propose leveraging the non-linear memory effect --- a subtle, persistent feature in the GW signal resulting from the cumulative impact of emitted gravitational waves --- as a novel approach to enhance parameter estimation accuracy. 
Through a comprehensive series of injection studies, encompassing both reduced and full parameter spaces, we evaluate the effectiveness of non-linear memory in various scenarios for aligned-spin systems. Our findings demonstrate the significant potential of non-linear memory in resolving the inclination-distance degeneracy, particularly for events with high signal-to-noise ratios (SNR $>$ 90) for the current generation of detectors or closer than 1 Gpc in the context of future detector sensitivities such as the planned LIGO A$^\sharp$ upgrade. The results also suggest that excluding non-linear memory from parameter estimation could introduce significant systematics in future LIGO A$^\sharp$ detections. This observation will hold even greater weight for next-generation detectors, highlighting the importance of including non-linear memory in GW models for achieving high-accuracy measurements for gravitational wave (GW) astronomy.

\end{abstract}

\maketitle

\section{Introduction}

Having observed nearly 100 gravitational wave events in recent years, and anticipating a multifold increase in detections during the current O4 run, gravitational wave detections open a new era for cosmology and astrophysics \cite{KAGRA:2021vkt, KAGRA:2021duu}. 
Many cosmological and astrophysical inferences that can be made from these GW detections rely on accurate measurements of the luminosity distance and inclination of the binary. Such inferences include constraining the Hubble constant through methodologies that involve the parameter estimation of luminosity distance from gravitational waves, which has the potential to solve the Hubble tension\cite{2021ApJ...909..218A, LIGOScientific:2021aug, Vasylyev:2020hgb, Schutz:1986gp, Nissanke:2013fka}. 
However, it is difficult to accurately measure the parameter of luminosity distance due to the degeneracy with the inclination angle, which is a common issue in the parameter estimation \cite{Nissanke:2009kt, Usman:2018imj, Chassande-Mottin:2019nnz}. Further, the difficulty in obtaining a precise measurement of the inclination angle due to the degeneracy also hinders our understanding of Gamma-ray bursts \cite{Clark:2014jpa, Goldstein:2017mmi}. 

One way to solve this degeneracy problem is by utilizing higher-order modes \cite{London:2017bcn}. While these modes are typically weak for symmetric mass ratio systems, they can be significant in asymmetric systems that involve a large black hole and a smaller counterpart. For binary systems that are undergoing precession, the changing inclination angle encodes more information in the waveform, breaking the degeneracy \cite{Vitale:2018wlg}. Additionally, for binary neutron stars which contain matter effects, the love relation can be employed to further break the degeneracy \cite{Xie:2022brn}.

However, in many of the LIGO-Virgo compact binary coalescence (CBC) events, the binaries do not have a high mass ratio, precession, or matter effect \cite{KAGRA:2021vkt}. As a result, we propose a complementary approach to tackle this issue for binary systems that are symmetric in mass and non-precessing, using the non-linear memory effect.
The non-linear memory (also referred to as null memory \cite{Bieri:2013ada}) effect is a permanent spacetime displacement phenomenon that can be interpreted as a gravitational wave component generated by the emission of gravitational waves itself \cite{Christodoulou:1991cr, Payne:1983rrr, Blanchet:1992br, Wiseman:1991ss, Thorne:1992sdb}. This effect is not an isolated phenomenon but a crucial element within the broader framework of gravitational wave physics. It stands as one of the pivotal aspects of the 'infrared triangle,' a conceptual model that intricately links gravitational-wave memory with BMS supertranslations and soft theorems, as explored in depth by Strominger \cite{Geroch:1981ut, Ludvigsen:1989kg, Strominger:2014pwa}. This interconnection underscores the fundamental nature of gravitational wave memory in understanding the underlying principles of gravitational wave emissions and their implications for theoretical physics. In this work, we limit ourselves to the so-called \textit{displacement memory} as it is the dominant memory effect which is available for the gravitational waves detectors. Other sub-dominant memory effects such as spin memory\cite{Nichols:2017rqr,Pasterski:2015tva} and center-of-mass memory \cite{Nichols:2018qac} are left for future investigations.  

Non-linear memory can be utilized to distinguish neutron star-black hole (NSBH) mergers from binary black hole (BBH) mergers for a certain part of the parameter space \cite{Tiwari:2021gfl}. While in the presence of matter effect, the non-linear memory effect also shows the ability to distinguish binary neutron star (BNS) from BBH and NSBH \cite{Lopez:2023aja}.

The morphology of the non-linear memory is influenced by various factors such as the total mass, mass ratio, and inclination. Predominantly residing in the (2,0) mode for non-precessing systems or the corresponding mode in the co-precessing frame for precessing binaries, the memory effect varies based on the orientation of the system -- maximized in edge-on and minimized for face-on systems \cite{Talbot:2018sgr}. This characteristic could serve as a potent tool for disentangling the inclination-distance degeneracy.

Despite its potential, the non-linear memory effect is typically weak and has not been found in the past LIGO-Virgo runs \cite{Hubner:2019sly,Hubner:2021amk}. The prospects of detection of non-linear memory from multiple events with LIGO-Virgo-KAGRA are discussed here \cite{Lasky:2016knh,Boersma:2020gxx,Grant:2022bla}. There are a few works to set the detection limit for memory in Einstein Telescope and space-based gravitational wave detectors \cite{Sun:2022pvh, Sun:2024nut, Goncharov:2023woe}. Additionally, \cite{Gasparotto:2023fcg} has demonstrated the potential of non-linear memory in resolving the inclination-distance degeneracy for the LISA with Fisher analysis.
Here we aim to investigate how non-linear memory will affect the parameter estimation results through a campaign of injection studies, and understand the parameter range where the inclusion of the memory effect in parameter estimation is necessary for ground-based detectors. We also aspire to understand if failing to consider the memory effect in parameter estimation could lead to systematics. 
In this work, we have studied the binary systems with aligned spins as generic spins can lead to complex features making it more challenging to study non-linear memory in isolation. Moreover, the non-linear memory peaks at equal mass systems where the effect of spin-precession is expected to be minimal.

This paper is organized as follows. In Section \ref{sec:nonlinear-memory}, we provide a brief explanation of the non-linear memory effect, including its properties. In Section \ref{sec:meth}, we briefly introduce the methods for parameter estimation, the waveform model, and the statistics used in this work. Finally, in Section \ref{sec:res}, we present our results obtained from the injection studies including the injections with reduced priors and full priors, as well as using the future LIGO-A$^\sharp$ upgraded sensitivity, and a selected real event detected by the LVK collaboration.

\section{Nonlinear memory }\label{sec:nonlinear-memory}

The non-linear memory component of gravitational waves can be calculated from the oscillatory modes $h^{\ell m}$ of gravitational waves for each spherical harmonic $_{-2}Y^{\ell m}$ using the equation \cite{Thorne:1992sdb,Talbot:2018sgr,Ebersold:2020zah,Mitman:2020bjf}

\begin{equation}\label{eq:mem}
    \begin{aligned}
h_{\mathrm{mem}}^{\ell m}= & \frac{R}{c} \sqrt{\frac{(\ell-2) !}{(\ell+2) !}} \sum_{\ell^{\prime}=2}^{\infty} \sum_{\ell^{\prime \prime}=2}^{\infty} \sum_{m^{\prime}=-\ell^{\prime}}^{\ell^{\prime}} \sum_{m^{\prime \prime}=-\ell^{\prime \prime}}^{\ell^{\prime \prime}} \\
& \times G_{m m^{\prime} m^{\prime \prime}}^{\ell \ell^{\prime} \ell^{\prime \prime} } \int_{-\infty}^{T_R} d t \dot{h}^{\ell^{\prime} m^{\prime}} \dot{\bar{h}}^{\ell^{\prime \prime} m^{\prime \prime}},
\end{aligned}
\end{equation}
where $R$ denotes the luminosity distance, $c$ represents the speed of light, and $T_R$ refers to the retarded time. The term $\dot{h}^{\ell m}$ is the time derivative of the strain, with the overbar indicating its complex conjugate. The coefficient $G_{m m^{\prime} m^{\prime \prime}}^{\ell \ell^{\prime} \ell^{\prime \prime}}$ is defined as follows:
\begin{equation}\label{eq:G}
    \begin{aligned}
& G_{m m^{\prime} m^{\prime \prime}}^{\ell \ell^{\prime} \ell^{\prime \prime}}=\int d \Omega^{\prime} \, \bar{Y}^{\ell m}\left(\Omega^{\prime}\right) _{-2}Y^{\ell^{\prime} m^{\prime}}\left(\Omega^{\prime}\right) _{-2}\bar{Y}^{\ell^{\prime \prime} m^{\prime \prime}}\left(\Omega^{\prime}\right) \\
& =(-1)^{m+m^{\prime \prime}} \sqrt{\frac{(2 \ell+1)\left(2 \ell^{\prime}+1\right)\left(2 \ell^{\prime \prime}+1\right)}{4 \pi}} \\
& \times\left(\begin{array}{ccc}
\ell & \ell^{\prime} & \ell^{\prime \prime} \\
0 & 2 & -2
\end{array}\right)\left(\begin{array}{ccc}
\ell & \ell^{\prime} & \ell^{\prime \prime} \\
-m & m^{\prime} & -m^{\prime \prime}
\end{array}\right). 
\end{aligned}
\end{equation}
It is an integral over three spin-weighted spherical harmonics, two with spin weight -2 and one ordinary spherical harmonic function with spin weight 0, which can be expressed in terms of Wigner 3-j symbols. The $_{-2}\bar{Y}^{\ell m}$ denote the complex conjugate of $_{-2}Y^{\ell m}$.

\begin{figure}[t]
    \centering
    \includegraphics[width=0.45\textwidth]{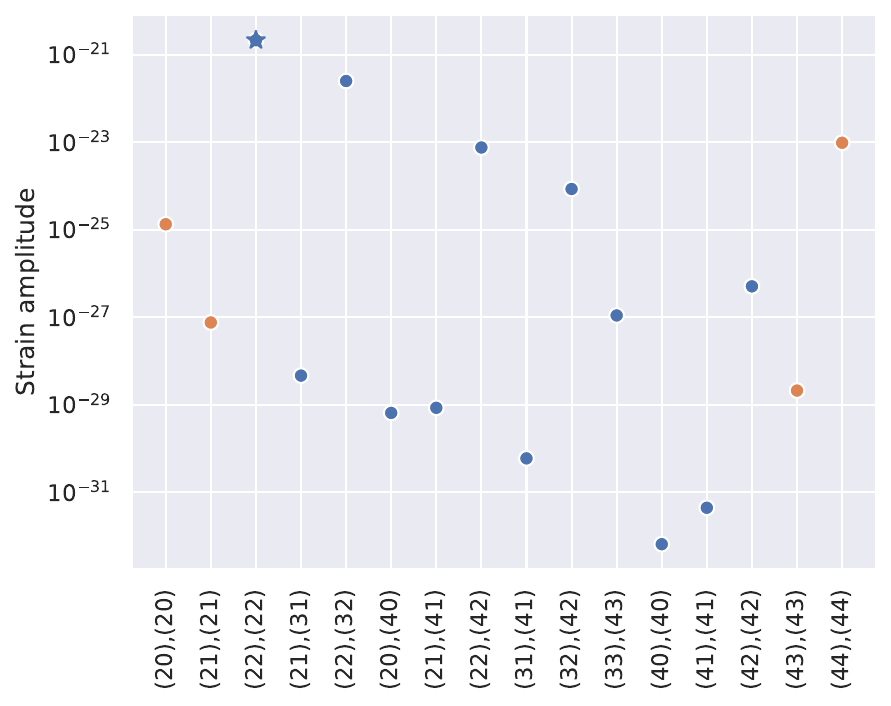}
    \caption{The amplitude of memory contribution for different $(\ell,m)$ mode combinations to $(2,0)$ mode. The blue points show the non-linear memory with positive amplitude while the red points show the ones with negative amplitude. The contribution from $(2,2)$ mode is shown in the star symbol.}
    \label{fig:hom_contribution}
\end{figure}

In equation \eqref{eq:mem}, we can see that the memory is a result of combinations of all $(\ell,m)$ modes. However, it should be noted that each combination contributes with a different amplitude due to the differences in the amplitude and phase of each mode, as well as the varying factor $G_{m m^{\prime} m^{\prime \prime}}^{\ell \ell^{\prime \prime}}$ in equation \eqref{eq:G}. The amplitude contributions of each combination are displayed in Fig.~\ref{fig:hom_contribution}. It is evident that the $(2,2)\times(2,2)$ combination is from 8 times to several orders of magnitude higher than the other modes. Considering the non-linear memory is already around two orders lower than the dominant $(2,2)$ mode, we can safely ignore the contribution from other modes to reduce the computational time significantly.

\begin{figure}[t]
    \centering
    \includegraphics[width=0.45\textwidth]{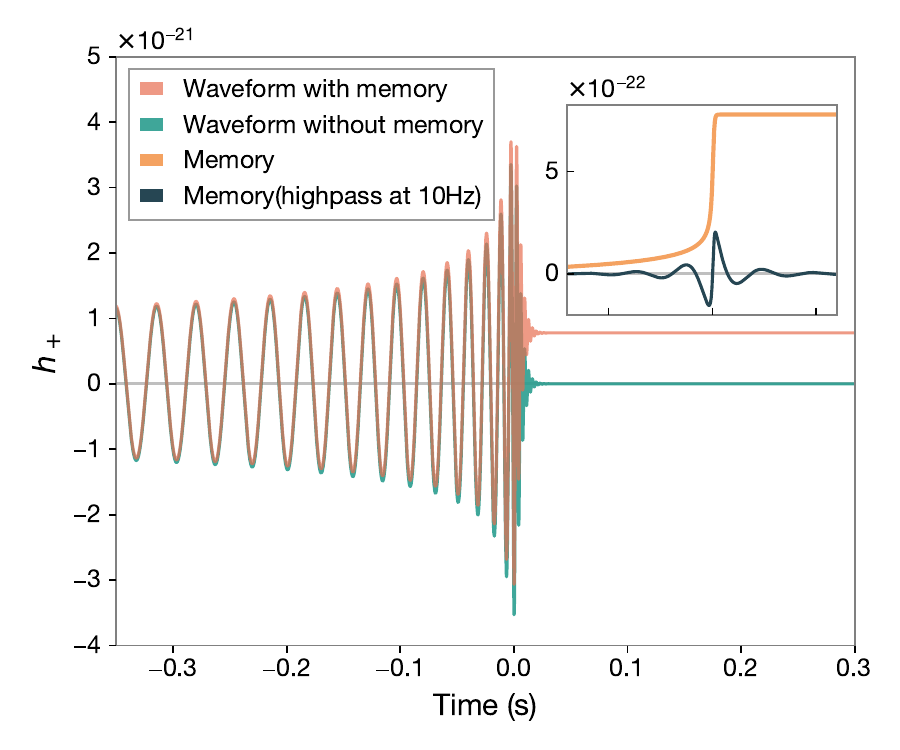}
    \caption{The waveforms shown in the main plot are the plus polarization of gravitational waves  (2,$\pm$2) oscillatory modes with and without their corresponding non-linear memory contribution to (2,0) modes. The inset shows the non-linear memory separately in orange and the same signal after a high pass at 10 Hz in dark blue, showing its nearly oscillatory behavior. These waveforms are generated using the waveform model described in Sec. \ref{sec:model:tmem} for a total mass 40 $M_\odot$, mass ratio 1, non-spinning, and edge-on binary black hole system at 100 Mpc.}
    \label{fig:memory_highpass}
\end{figure}

Non-linear memory contributions alone exhibit characteristics similar to step functions. They increase monotonically and slowly with the compact binary getting closer, then sharply rise at merger, and finally induce a permanent amplitude shift. After a merger event, memory appears as a flat line, indicating its nearly zero frequency as shown in the orange line of the inset in Fig.~\ref{fig:memory_highpass}. Ground-based detectors, which are sensitive to tens to a few thousand Hz, cannot detect this ``0 Hz'' permanent displacement directly. To truly observe the memory effect, one would need freely-falling observers, indicating a permanent, net displacement. This necessitates the application of a high-pass filter to the waveform as shown in the deep blue line in the inset of Fig.~\ref{fig:memory_highpass}. The plus polarization waveform contains only (2,$\pm$2) oscillatory modes with/without non-linear memory in $(2,0)$ are shown in Fig.~\ref{fig:memory_highpass} with red and green respectively.

\begin{figure}[t]\label{fig:snr}
    \centering
    \includegraphics[width=0.5\textwidth]{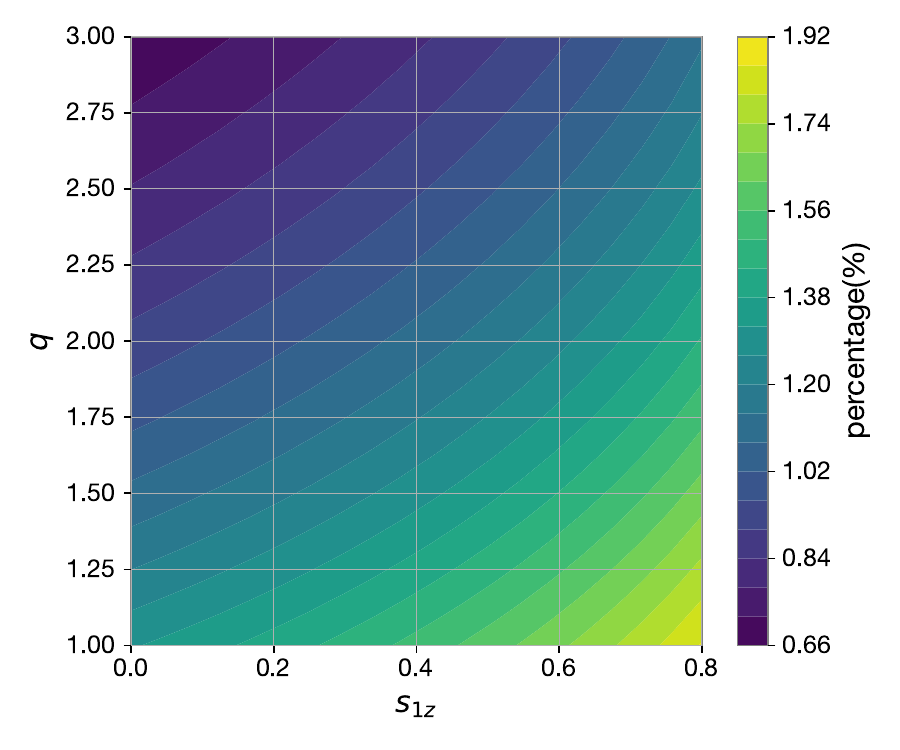}
    \caption{The SNR relative fraction of the real part of non-linear memory contribution in (2,0) mode and dominant (2,2) mode in different aligned spin by varying the primary black hole spin in the z-direction, $s_{1z}$, and mass ratio $q$. Here we have used the zero-detuned high-power LIGO noise power spectral density (PSD) as a representative of the current generation of ground-based detectors.  
    }
    \label{fig:snr}
\end{figure}

Fig.~\ref{fig:snr} shows the SNR relative fraction of the plus polarization of non-linear memory contribution and the dominant (2,2) oscillatory mode for different aligned spin and mass ratios.
We see that the non-linear memory has a much higher SNR relative fraction with respect to the dominant (2,2) mode for a symmetric mass ratio system. This property makes non-linear memory a complementary method for constraining the distance of a gravitational wave source for systems when higher-order modes are negligible, such as systems with a symmetric mass ratio.

\section{Methodology}\label{sec:meth}

\subsection{Bayes parameter estimation}

In this study, the parameters we are exploring are highly degenerate, which cannot be accurately represented by the Fisher matrix. Therefore, we have opted to use Bayes inference. This method can be summarized by the following simplified equation \cite{Thrane:2018qnx}:

$$\underbrace{P(\mathrm{parameter}| \mathrm{data})}_{\text{posterior}} \propto \underbrace{P(\mathrm{data}| \mathrm{parameter})}_{\text{likelihood}} \times \underbrace{P(\mathrm{parameter})}_{\text{prior}}\,,$$

The posterior is the probability of parameters with the given data, which is proportional to the likelihood times the prior. The likelihood is evaluated with the data and waveform model. In this case, we utilize the \texttt{IMRPhenomT} \cite{Estelles:2020osj} waveform model with memory contribution as it is both computationally efficient and sufficiently accurate for our purposes. More details about this model can be found in the following section. To thoroughly study the effect of non-linear memory with limited computational resources, we first apply a simple prior that samples only inclination, distance, polarization, and phase parameters and keeps other parameters as delta functions. Then we open all other parameters apart from the non-aligned spins for a more complete and realistic study.

The software we are using for this study is \texttt{parallel-bilby} \cite{bilby, pbilby} with the \texttt{dynesty} \cite{dynesty} sampler, which allows us to run the parameter estimation in parallel and significantly reduce the run time.

For the injected signal, we use \texttt{NRHybSur3dq8\_CCE} waveform model, which is the most complete waveform model containing both the oscillatory and non-linear memory component of (2,0) mode derived from numerical relativity \cite{Yoo:2023spi}. 
In this study, we only use the (2,2) mode and (2,0) mode from \texttt{NRHybSur3dq8\_CCE}.
The generated waveforms are injected into both zero-noise and colored Gaussian noise with selected sensitivity curves for ground-based detectors. Explorations with zero noise can be seen as the average of all noise realizations and help us understand the intrinsic degeneracies present, while by introducing colored noise, we can gain insight into the impact of noise on such a weak signature.

In order to calculate memory accurately from our waveform model, we need to integrate the oscillatory modes precisely. This requires a high sample rate to reduce the error in the integration. However, a higher sample rate can significantly slow down waveform generation. Therefore, we use a different sample rate for each different total mass as shown in Table \ref{tb:sample_rate}.

\begin{table}[t]
\caption{\label{tb:sample_rate} Sample rates for different total mass range.
}
\begin{ruledtabular}
\begin{tabular}{@{}cl}
Total mass & Sample rate \\
\colrule
$> 80 M_\odot$ & 1024 Hz \\
$(55 M_\odot, 80 M_\odot) $ &  2048 Hz \\
$(25 M_\odot, 55 M_\odot) $ &  4096 Hz \\
$(15 M_\odot, 25 M_\odot) $ &  8192 Hz \\
$< 15 M_\odot$ & 16384 Hz
\end{tabular}
\end{ruledtabular}
\end{table}

\subsection{Aligned spin waveform model with memory}\label{sec:model:tmem}

To calculate the memory contribution using equations \eqref{eq:mem} and \eqref{eq:G}, we require a time-domain waveform model to perform the time integral. We have selected \texttt{IMRPhenomT} as our baseline model. This model is fast and accurate, making it an ideal choice for our needs. Since the contribution of other modes to memory is much smaller than that of the $(2, 2)$ mode (See Fig.~\ref{fig:hom_contribution}), we can discard them in the integral to reduce computational cost. Then equation \eqref{eq:mem} and \eqref{eq:G} reduces to 

\begin{equation}
h_{\mathrm{mem}}^{2 0}= \frac{R}{c} \frac{1}{\sqrt{24}} \sum_{m^{\prime}=-2,2} G_{0 m^{\prime} m^{\prime}}^{2 2 2} \int_{-\infty}^{T_R} d t \dot{h}^{2 m^{\prime}} \dot{\bar{h}}^{2 m^{\prime}}.
\end{equation}

From this equation, the non-linear memory can be calculated by taking the derivative of $(2,\pm 2)$ oscillatory modes from \texttt{IMRPhenomT} and performing the integral over time. 
The gravitational wave strain can then be calculated by inserting the memory contribution to the $(2,0)$ mode together with the oscillatory $(2, \pm 2)$ modes in ~\cite{Thorne:1980ru},

\begin{equation}
    h\left(\iota,\phi_0\right)
    = \sum_{\ell=2}^{\infty} \sum_{m=-\ell}^{\ell} 
    h_{\ell m}
    \,_{-2}Y^{\ell m}\left(\iota,\phi_0\right).
\end{equation}

\begin{figure}[t]
    \centering
    \includegraphics[width=0.48\textwidth]{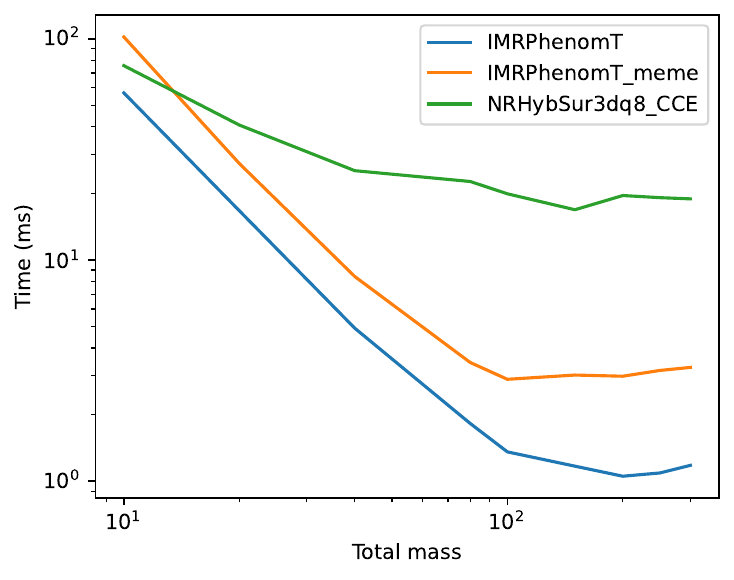}
    \caption{Comparison of computational performance of \texttt{IMRPhenomT}, \texttt{IMRPhenomT} with non-linear memory (labelled as \texttt{IMRPhenomT\_meme}) and \texttt{NRHybSur3dq8\_CCE}. The timing data is averaged over 500 runs for each total mass.}
    \label{fig:wf_timing}
\end{figure}

\begin{figure}[t]
    \centering
    \includegraphics[width=0.48\textwidth]{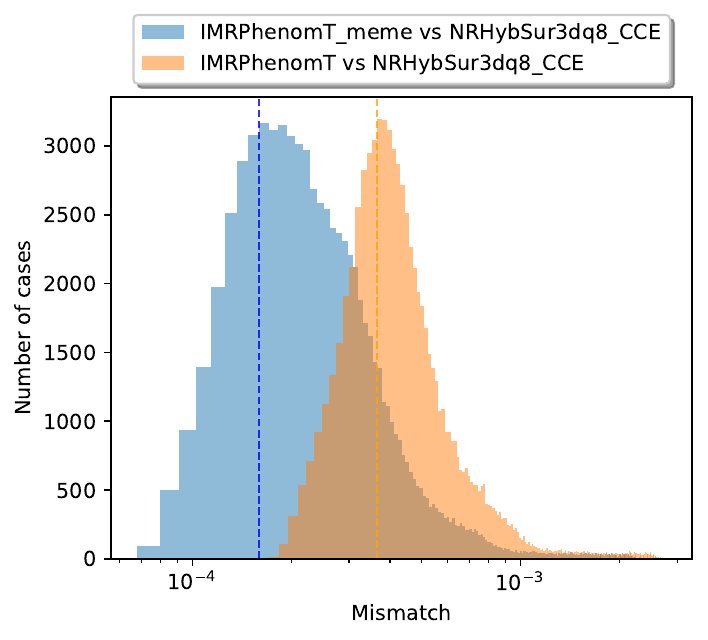}
    \caption{The mismatch plot between \texttt{IMRPhenomT} and \texttt{NRHybSur3dq8\_CCE} (shown in blue), \texttt{IMRPhenomT\_meme} and \texttt{NRHybSur3dq8\_CCE} (shown in orange). The peaks for each mismatch distribution are indicated by dashed lines.}
    \label{fig:wf_mismatch}
\end{figure}

Fig.~\ref{fig:wf_timing} compares the computational performance of various waveform models as a function of total masses. It is observed that the \texttt{NRHybSur3dq8\_CCE} model takes a few to ten times longer to perform compared to the \texttt{IMRPhenomT\_meme} model. Due to this significant difference, the \texttt{NRHybSur3dq8\_CCE} model is less practical for systematic PE study. Hence, we have chosen to use the \texttt{IMRPhenomT\_meme} model for this exploration.

The figure displayed in Fig.~\ref{fig:wf_mismatch} illustrates the mismatch between the waveforms generated by \texttt{IMRPhenomT} and \texttt{IMRPhenomT\_meme} as compared to those generated by \texttt{NRHybSur3dq8\_CCE}. The mismatch for two waveform $h_1$ and $h_2$ is defined as
\begin{equation}
    \textrm{mismatch}(h_1, h_2) = 1 - \max_{\phi, t, \psi} O(h_1, h_2)
\end{equation}
which maximizing the overlap $O(h_1, h_2)$ over polarization angles $\psi$, phase $\phi$, and time shifts $t$. The overlap $O(h_1, h_2)$ is given by
\begin{equation}
    O(h_1, h_2) = \frac{(h_1 \mid h_2)}{\sqrt{(h_1 \mid h_1)(h_2 \mid h_2)}}
\end{equation}
and the power spectrum $S(f)$ weighted inner product $(a \mid b)$ is defined as
\begin{equation}
(a \mid b):=4 \operatorname{Re} \int_0^{f_{\max }} \frac{\tilde{a}(f) \tilde{b}(f)^{\star}}{S(f)} \mathrm{d} f
\end{equation}
We generate the waveforms for each waveform model only with $(2,2)$ and $(2,0)$ modes, and compute the mismatches for binary black holes with total mass 20-80 $M_\odot$, mass ratio from 1 to 3, and $s_{1z}$ from 0 to 0.8. The histogram peak indicates a factor of 2 improvement in mismatches for \texttt{IMRPhenomT\_meme} vs \texttt{IMRPhenomT} with \texttt{NRHybSur3dq8\_CCE}. From this mismatch study, we can see that both \texttt{IMRPhenomT\_meme} and the underlying base model \texttt{IMRPhenomT} are highly accurate and sufficient for use in this study. While the relatively small improvement due to the inclusion of the memory effect means it may be difficult to conclusively claim detection of the memory effect at current detector sensitivities, the complementary phenomenology of the memory effect relative to the oscillatory signal makes it perfect for assisting in breaking the distance-inclination degeneracy.

\subsection{Relative fractional improvement}

To quantify the relative improvement of the measurement for the luminosity distance to the source, we introduce the relative fractional improvement inspired by \cite{Xie:2022brn},
\begin{align}\label{eq:delta_dL}
    \Delta D_L &= 100\% \times \left[1 - \frac{(\delta D_L)_\text{mem}}{(\delta D_L)_\text{nomem}}\right], 
\end{align}
where $(\delta D_L)_\text{mem}$ represents the 1-$\sigma$ credible interval of luminosity distance posterior using the non-linear memory waveform model. $(\delta D_L)_\text{nomem}$ is the corresponding value obtained using the waveform without non-linear memory. A better constraint on distance in the posterior will give a higher $\Delta D_L$. However, this is insufficient to show if the recovered distances are close to the injected values. Thus, we will also use the mean-squared error (MSE) of the $1-\sigma$ luminosity distance posterior versus the injected value together with $\Delta D_L$ to quantify the improvement. We define a quantity $\Delta \text{MSE}$ in the same way,
\begin{align}\label{eq:delta_MSE}
    \Delta \text{MSE} &= 100\% \times \left[1 - \frac{ \text{MSE}_\text{mem}}{\text{MSE}_\text{nomem}}\right],
\end{align}
where
\begin{equation}
    \text{MSE} = \frac{1}{n} \sum_{i=1}^{n} \left(D_L^i - \hat{D}_L\right)^2.
\end{equation}
Here $\hat{D}_L$ is the injected true value of luminosity distance.

\subsection{(2,0) mode SNR}
Non-linear memory can be considered as one of the higher-order mode multipoles. Thus we can apply the conventional higher-order mode multipole SNR to set out theoretical expectations.
Inspired by \cite{Mills:2020thr}, we proposed a new quantity memory SNR $\rho_{20}$ , we can define it as 
\begin{align}
    \rho_{20}= \rho_{22}^{+} \alpha_{20} R_{20}^+,
\end{align}
where $\rho_{22}^{+}$ is the SNR of $(2,2)$ mode plus polarization and
\begin{align}
    R^+_{20} & = \frac{\sin^2 \iota}{1 + \cos^2 \iota},\\
    \alpha_{20}  & = \frac{\sigma_{20}}{\sigma^+_{22}},
\end{align}
while the sensitivity of the detector to the $(\ell,m)$ harmonic, $\sigma_{\ell m}$, is given by
\begin{align}
    \sigma_{\ell m} = \sqrt{(\tilde{h}_{\ell m}|\tilde{h}_{\ell m})}.
\end{align}
where $\tilde{h}_{\ell m}$ is the higher order harmonics $(\ell, m)$ waveform in the frequency domain

The orthogonal component of non-linear memory contribution with dominant 22 multipole is given by
\begin{align}\label{eq:snr20}
    \rho_{20}^{\perp} = \rho_{20} \sqrt{1 - O(h_{20},h_{22})^2},
\end{align}
where $O(h_{20},h_{22})$ is the overlap of the modes (2,0) and (2,2).

\section{Results}\label{sec:res}
We present a series of injection studies to investigate the effectiveness of non-linear memory in breaking the degeneracy of distance and inclination with a reduced prior, using LIGO-Virgo O5 sensitivity \cite{sensitivity_curve_o5}.
We then vary all the priors for the non-precessing system to explore the improvement in a more realistic situation. Finally, we discuss the potential improvement that non-linear memory can bring to the distance constraint for the future LIGO A$^\sharp$.

\subsection{Degeneracy breaking for reduced prior}\label{sec:res:reduced_prior}
\begin{figure}[t]
    \centering
    \includegraphics[width=0.48\textwidth]{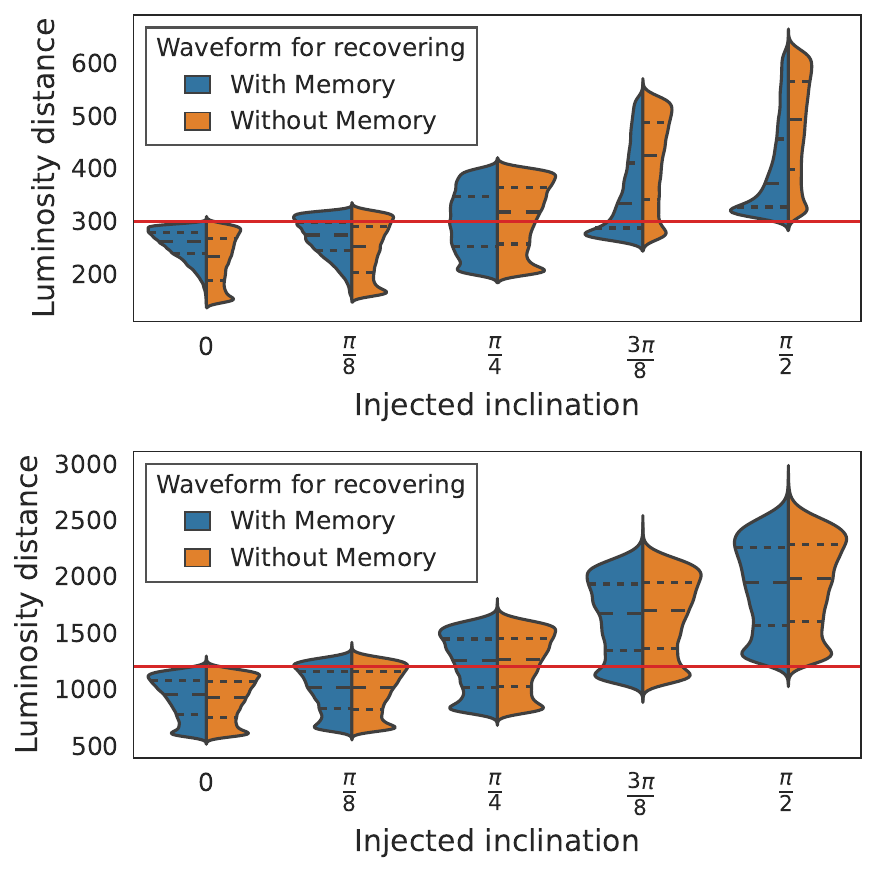}
    \caption{The upper and lower panels of the violin plots show the luminosity distance from 300 Mpc and 1200 Mpc injection and reduced prior, respectively. The blue posteriors are obtained using a waveform model with non-linear memory, while the orange ones are from the recovery with IMRPhenomT waveform model. The quartile lines of credible intervals are represented by the dashed lines in the violin plot.}
    \label{fig:violin_4p}
\end{figure}

\begin{figure*}[t]
    \subfloat[\label{fig:dL_iterval_4p}]{%
    \includegraphics[width=0.325\textwidth]{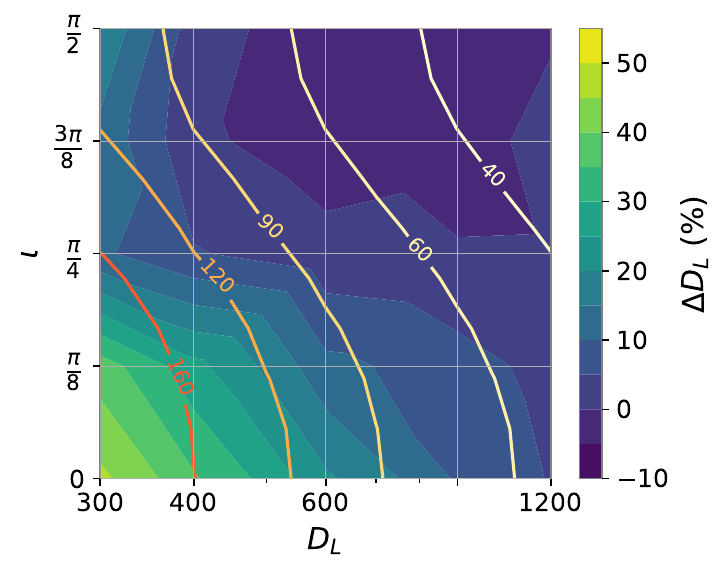}
    }\hfill
    \subfloat[\label{fig:MSE_4p}]{%
    \includegraphics[width=0.325\textwidth]{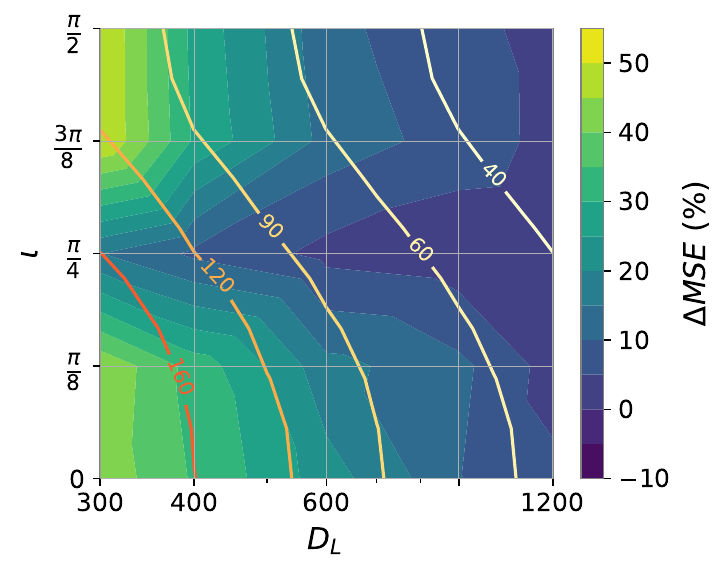}
    }\hfill
    \subfloat[\label{fig:q_dL}]{%
    \includegraphics[width=0.325\textwidth]{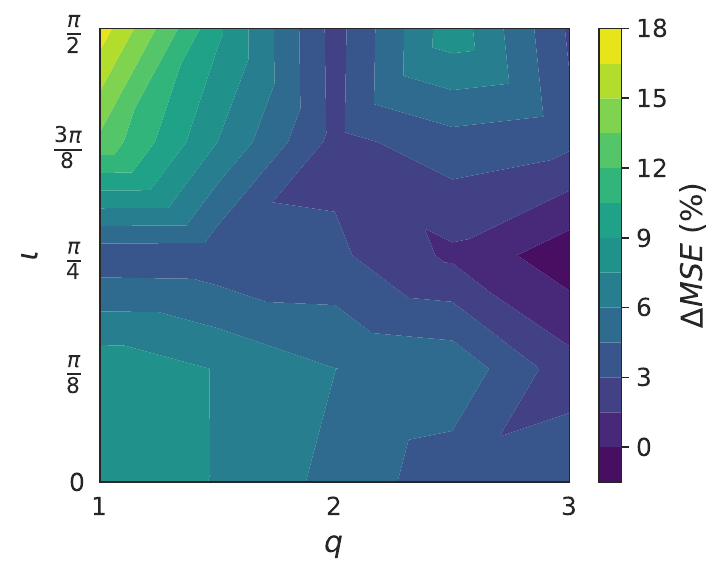}
    }\hfill
    \caption{Panel (a) and (b) show the relative fractional improvement of luminosity distance $\Delta D_L$ and $\Delta \mathrm{MSE}$ with reduced prior. 25 injections were performed at 5 luminosity distances and 5 inclination angles. The contour color represents the relative improvement of the luminosity distance $d_L$ (defined in Eq. \eqref{eq:delta_dL} and \eqref{eq:delta_MSE}). The yellow lines represent the SNR contour of the injected signal. Panel (c) shows the $\Delta \mathrm{MSE}$ with reduced prior of varying mass ratio. The results align with what we see in the SNR plot in Fig \ref{fig:snr}, that higher mass ratio $q$ leads to less gain in measuring the distance from introducing non-linear memory.}
\end{figure*}

To explore the ability to break the degeneracy of inclination and distance with non-linear memory, we first use a reduced prior which only varies 3 parameters (inclination $\iota$, luminosity distance $D_L$, and phase $\phi$) while keeping other parameters as a delta function for the injected parameters. 
This approach allows us to understand better how including the non-linear memory effect impacts the measurement of inclination and distance alone, without the need to account for additional parameters. Additionally, it reduces the computational burden of the analysis, enabling us to explore a broader parameter space, including variations in mass ratio.

We first inject 25 sets of parameters with 5 distances (300, 400, 600, 900, and 1200 Mpc) and 5 inclination angles (evenly-spaced values in the range of 0 to $\pi/2$). The mass ratio is set to 1 and the spins are set to 0. The total mass is 40 $M_\odot$.

In Fig.~\ref{fig:violin_4p}, we can see the posteriors of two different injection distances, 300 Mpc and 1200 Mpc, in violin plots. These plots demonstrate the enhancement of distance measurement using non-linear memory for high (around 160 at 300 Mpc) and low SNR (around 40 at 1200 Mpc). The upper panel of the plots represents the high SNR case, injected at 300 Mpc, where we can observe that the luminosity distance constraint is much better for face-on and edge-on systems. Additionally, for all inclination angles, the constraint peaks at the injected distance. On the other hand, for the low SNR case, injected at 1200 Mpc, the improvement is not as obvious, but the quartiles represented by the dashed line in the violin plots still show that the recoveries with a non-linear memory model provide better recovery of the injected value.

To quantify the relative fractional improvement of the luminosity distance measurement, we plot the $\Delta D_L$ and $\Delta \mathrm{MSE}$ in Fig.~\ref{fig:dL_iterval_4p} and \ref{fig:MSE_4p}. From Fig.~\ref{fig:dL_iterval_4p}, we can see that the highest improvement is from the closer distance and the face-on system, rather than the edge-on system, where the non-linear memory amplitude is highest, as we might expect.
This can be explained by the SNR contour plot overlaid on the $\Delta D_L$ contour, which shows the improvement on the $\Delta D_L$ following the SNR increase. The relative improvement on the width of the 90\% credible interval only provides information about the constraint, but it can't show the accuracy of the posterior. Therefore, we have also calculated the MSE and plotted the relative fractional improvement of MSE defined in Eq. \eqref{eq:delta_MSE} in Fig.~\ref{fig:MSE_4p}. We observe that even though the SNR at the same distance for the edge-on system is lower than that of the face-on system, due to the maximization of non-linear memory, the $\Delta \mathrm{MSE}$ is good for both systems.

In addition to the inclination-distance grid injections, we also performed 25 injections with varying mass ratios and inclination, of non-linear memory. The injected masses are all set to 40 $M_\odot$ and spins are set to 0. Here, we choose to fix the SNR to 60 instead of injecting them at the same distance. We also open the mass ratio in the priors for these parameter estimation runs. The results of $\Delta \mathrm{MSE}$ of luminosity distance are plotted in Fig.~\ref{fig:q_dL}. We see that the improvement of distance constraint follows what we expect. The more asymmetric mass ratio a binary black hole system has, the worse the improvement in the constraint.

We further explored injections with similar parameters, varying the aligned spin magnitude $s_{1z}$, as detailed in Appendix \ref{sec:align_spin}. The findings indicate that when the spin magnitude $s_{1z}$ is less than 0.6, the enhancements attributable to memory effects remain largely consistent. Consequently, our focus on cases with zero spin is representative of most scenarios involving aligned spin. 

\subsection{Full parameter recovery}

\begin{figure*}[t]
    \subfloat[\label{fig:mse_full}]{%
    \includegraphics[width=0.36\textwidth]{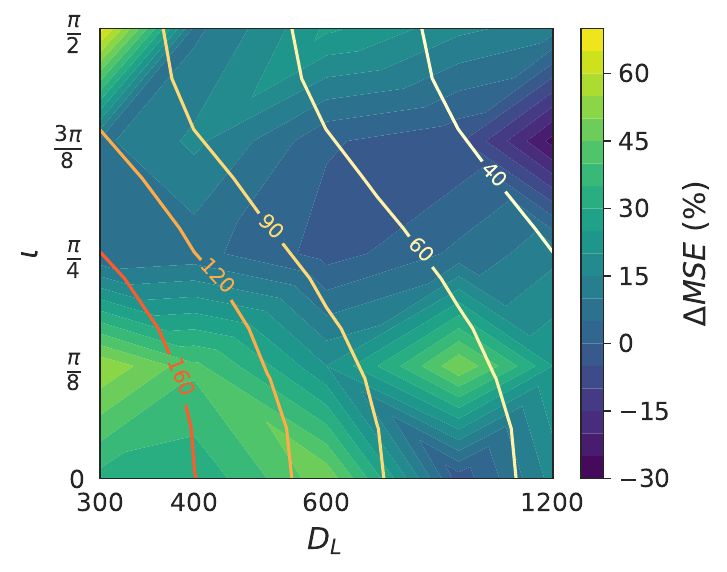}
    }
    \subfloat[\label{fig:contour_full}]{%
    \includegraphics[width=0.34\textwidth]{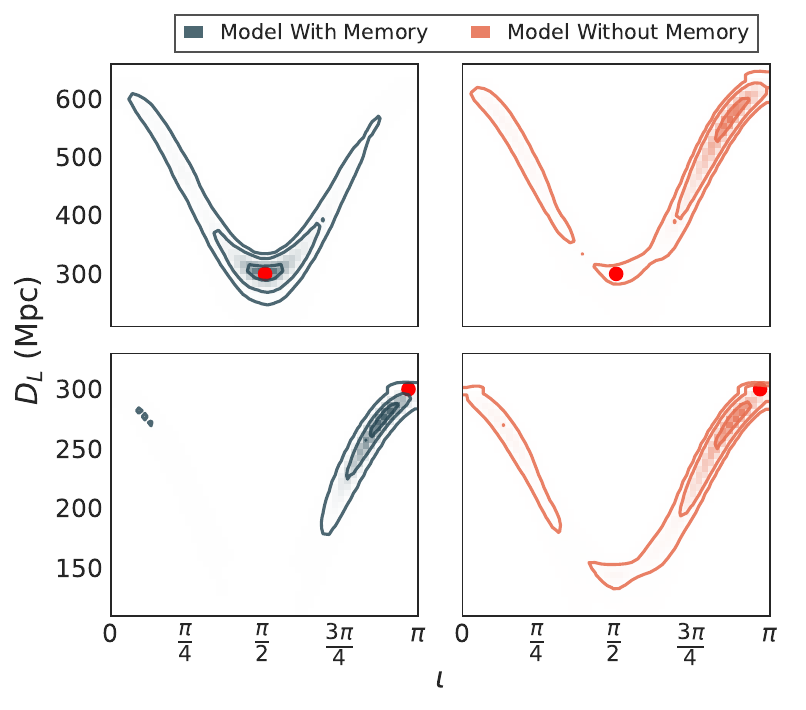}
    }
    \hfill
    \caption{Panel (a) shows the relative fractional improvement of MSE ($\Delta \mathrm{MSE}$ defined in \eqref{eq:delta_MSE}) with full aligned spin prior. 25 injections are performed with 5 luminosity distances and 5 inclination angles (evenly distributed from 0 to $\pi/2$). The yellow lines are the SNR contour of the injected signal. Panel (b) displays selected contour plots depicting the inclination and luminosity distance from the full parameter injections. The blue contours show the posteriors using the non-linear memory waveforms, while the orange contours in the same row show the corresponding ones without. The red dots in the plots indicate the injected values. }
\end{figure*}

In the case of performing PE on a real signal, we won't be able to know the other parameters such as total mass and mass ratio in prior. Therefore, we performed another Bayes inference exploration with full priors for aligned-spin binary black holes. 

We injected the same 25 sets of parameters with varied distances and inclinations as the previously reduced prior section. The contour plot of $\Delta \mathrm{MSE}$ is shown in Fig.~\ref{fig:mse_full}. Compared with the same contour plot in Fig.~\ref{fig:MSE_4p}, we can notice that with more parameters open for sampling in the prior, the improvement of the mean-squared error drops for the low SNR scenarios. For the high SNR cases, taken to be, when SNR is greater than 90, the $\Delta \mathrm{MSE}$ shows a similar amount of improvement. We can also notice that in the contour plot near inclination $\pi/3$ and distance $D_L$ around 1200 Mpc give worse MSE than the posteriors from the recovery without memory below SNR 40. This can be attributed to the very low (but not negligible) SNR available in the injection for the (2,0) mode which makes the posterior broader when sampling with memory. Also, the contribution from the weak oscillatory part of the (2,0) which is not available in the sampling model, can lead to further worsening of the posterior. This feature is resolved for the edge-on system (inclination of $\pi/2$) as the (2,0) mode will have more SNR. 

In Fig.~\ref{fig:contour_full}, we picked two data points from the injections, which are the edge-on ($\iota = \pi / 2$) and face-on ($\iota = 0 \textrm{ or } \pi$) system in the luminosity distance $D_L = 300$ Mpc. The contour plots from using the non-linear memory model are shown in the left panels in blue and the corresponding ones without non-linear memory are shown in the right panel in orange. We see that for the edge-on system, the posterior from using the non-linear memory waveform is constrained well around the injected value, while for the waveform model without non-linear memory, the recovered values are totally off. For the face-on system, both waveform models peak near the injected value. However, using non-linear memory can significantly reduce the bimodality, enable us to distinguish a face-on system and an edge-on system, and give also much better constraint on the distance.

These results give us hints that for low SNR (SNR $<$ 90), adopting non-linear memory in the parameter estimation won't gain too much improvement but will have a higher cost on the computing resources with adding additional mode for LIGO O5. 
For the high SNR events, ignoring the non-linear memory effect will introduce significant systematics in the distance-inclination measurement. 
The $\rho_{20}^{\perp}$ of the injections is greater than 2.3 for the ones with a luminosity distance closer than 550 Mpc. It is corresponding to SNR 90 when the inclination is $\pi/4$.
For current detectors, it is not likely to have such a high SNR and close-distance event. Thus in the next section, we will continue the exploration with the proposed LIGO A$^\sharp$ sensitivity curve \cite{sensitivity_curve_asharp}.

\subsection{Case of LIGO A$^\sharp$}

\begin{figure*}[t]
    \subfloat[\label{fig:mse_full_asharp}]{%
        \includegraphics[width=0.36\textwidth]{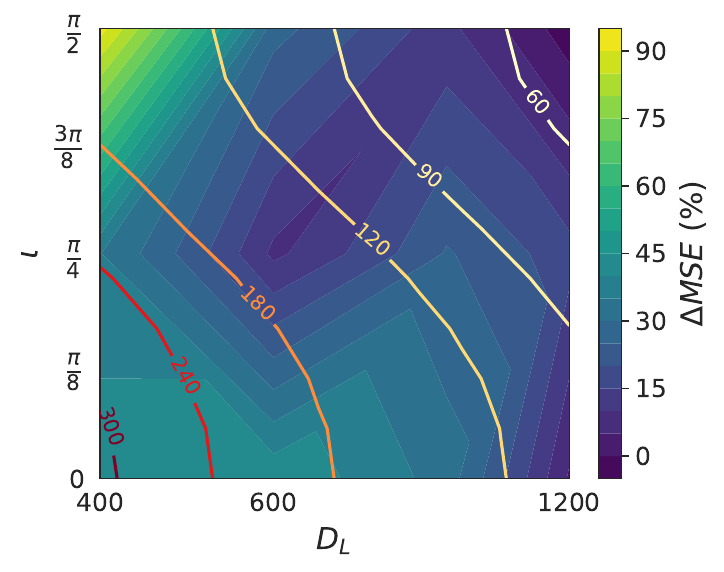}
    }\hfill
    \subfloat[\label{fig:contour_o5_asharp_600}]{%
        \includegraphics[width=0.31\textwidth]{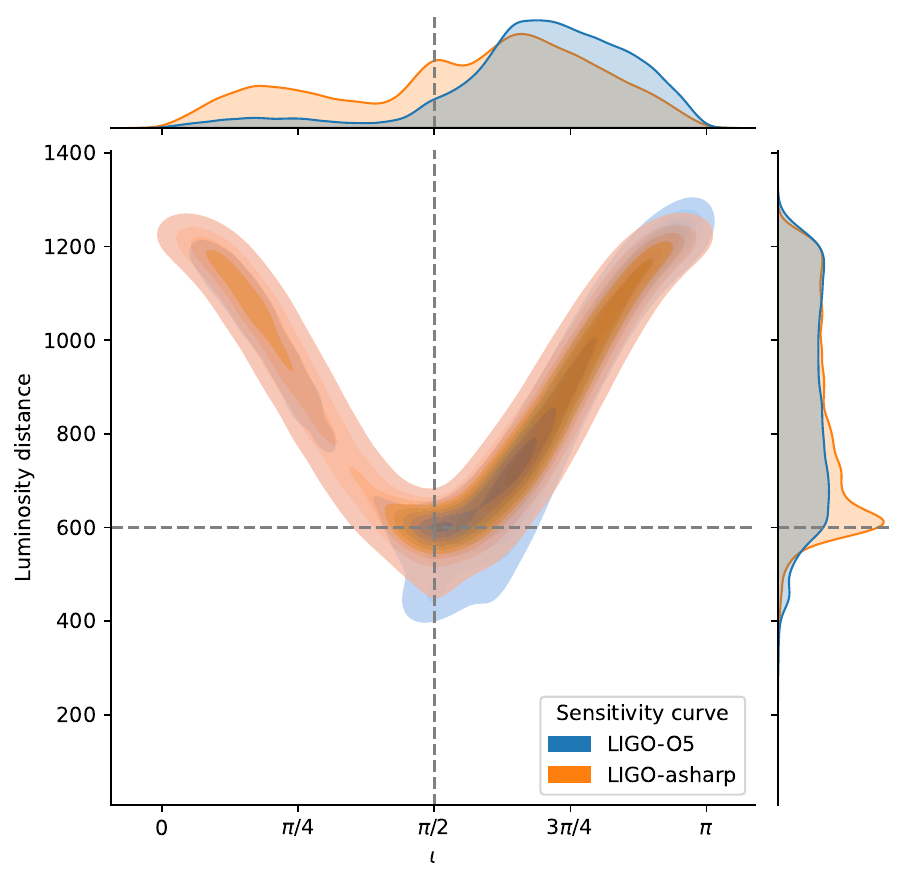}
    }\hfill
    \subfloat[\label{fig:contour_o5_asharp_400}]{%
      \includegraphics[width=0.31\textwidth]{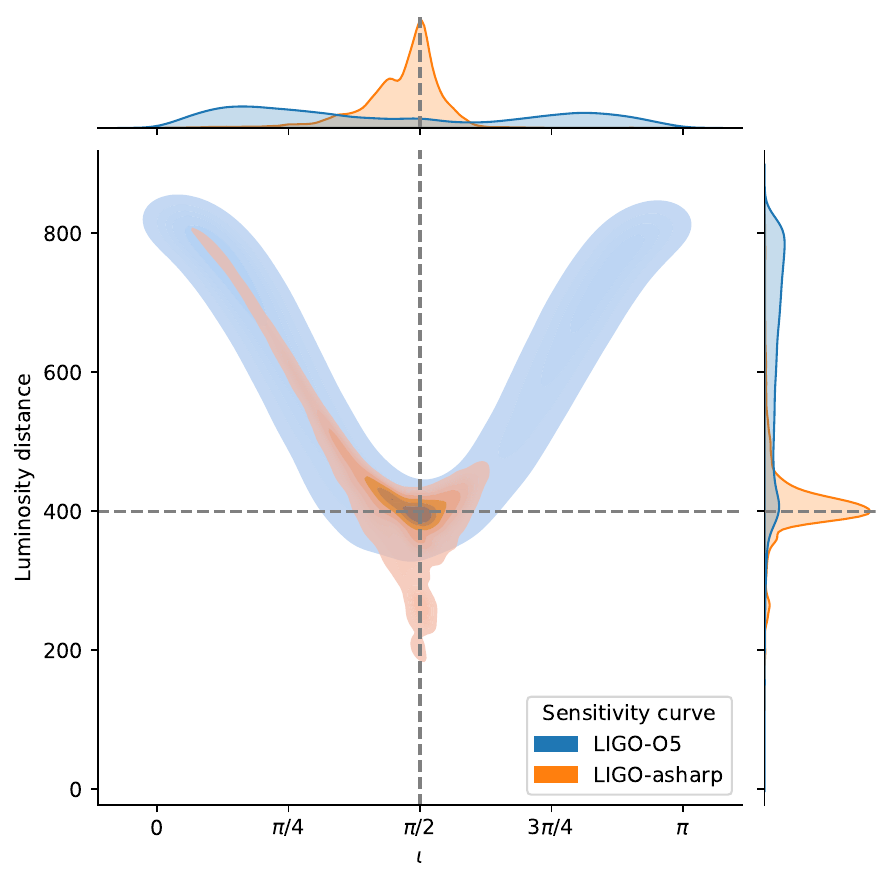}
    }
     \caption{Panel (a) shows Relative fractional improvement of MSE ($\Delta \mathrm{MSE}$) for full aligned-spin prior, with A$^\sharp$ sensitivity curve. 15 injections are performed with 5 luminosity distances and 3 inclination angles (0, $\pi/4$, and $\pi/2$). The yellow lines are the SNR contour of the injected signal. Panel (b) and (c) show contour plots of select results with LIGO O5 and LIGO A$^\sharp$ sensitivity curves at the same distance ($D_L = 400/600$ Mpc) and inclination ($\iota = \pi / 2$) in the same plot. The dashed lines indicate the injected values. }
\end{figure*}

From the exploration in the previous sections, we see that to break the degeneracy between inclination and luminosity distance, a high SNR greater than 90 is required. This means that for LIGO O5 sensitivities the distance should be closer than 300 - 600 Mpc, depending on the total mass and inclination angle. These kinds of close events are very few seen in O3 \cite{KAGRA:2021vkt}. Therefore, we continue our exploration with the future upgrade of LIGO A$^\sharp$ \cite{a-sharp}. Using the sensitivity curve of A$^\sharp$ is not simply equivalent to increasing the SNR of the overall signal, since the A$^\sharp$ will also increase the sensitivity in the low frequency more, where most of the energy of memory lives. 

We performed similar injection sets for parameters as in the previous section. Due to the limited computational resources, we reduced the injection parameter space to 15 sets, with 3 inclination angles (0, $\pi/4$, and $\pi/2$) for 5 different distances. The injected distance started from 400 Mpc because the sensitivity of A$^\sharp$ is much higher. The SNR below a distance of 400 Mpc is too high, which will cost too much time to finish.

The $\Delta \mathrm{MSE}$ are plotted in Fig.~\ref{fig:mse_full_asharp}. This graph displays the improvement of MSE up to 90\%. Even at around 1 Gpc, $\Delta \mathrm{MSE}$ can get around 10\% - 20\% of improvement. This means at that sensitivity, ignoring the non-linear memory effect will introduce significant systematics in the distance measurement for most of the significant events.
This is estimated to affect nearly 30\% of events reported in the GWTC-3 catalogue \cite{KAGRA:2021vkt} assuming they were detected by LIGO-A$^\sharp$, taking the mean measured distance as the ``true'' distance to the source.

We also plotted the comparison of inclination-distance posteriors for 2 select injection sets in both O5 and A$^\sharp$ sensitivity curve in Fig.~\ref{fig:contour_o5_asharp_600} and \ref{fig:contour_o5_asharp_400}. Fig.~\ref{fig:contour_o5_asharp_600} shows the inclination-distance contour for the edge-on system at 600 Mpc. We can see that under A$^\sharp$ sensitivity, the distance peaks at the injected value. Although the result using O5 sensitivity shows improvement in the MSE compared to the model without non-linear memory Fig.~\ref{fig:mse_full}, it is still largely unconstraint. 

Fig.~\ref{fig:contour_o5_asharp_400} shows the inclination-distance contour for the edge-on system at 400 Mpc. For the O5 sensitivity, the SNR is around 80 at this distance. This comparable low SNR provides an improvement in the distance measurement seen from \ref{fig:mse_full}, but still not enough to get rid of the bimodality. However, for A$^\sharp$ the SNR is 150, and both inclination and distance are nicely peaked at the injected values.

To conclude, for LIGO A$^\sharp$, we will need memory to resolve distance-inclination degeneracy 
since we anticipate that the majority of significant events will have a distance closer than 1 Gpc, based on estimates using the GWTC-3 catalogue.
Besides, for LIGO A$^\sharp$, the $\rho_{20}^{\perp} > 2.7$ when the distance is closer than 1 Gpc for total mass 40 $M_\odot$ binaries. For the same system at a distance of 600 Mpc, the $\rho_{20}^{\perp}$ is around 4.5.
For now, events below a distance of 600 Mpc account for 12\% of our total events in the GWTC-3 catalogue taking the mean measured distance, and those below 1 Gpc account for 29\%.

\subsection{A test on real data}

\begin{figure*}[t]
    \subfloat[\label{fig:joint_real}]{%
        \includegraphics[width=0.36\textwidth]{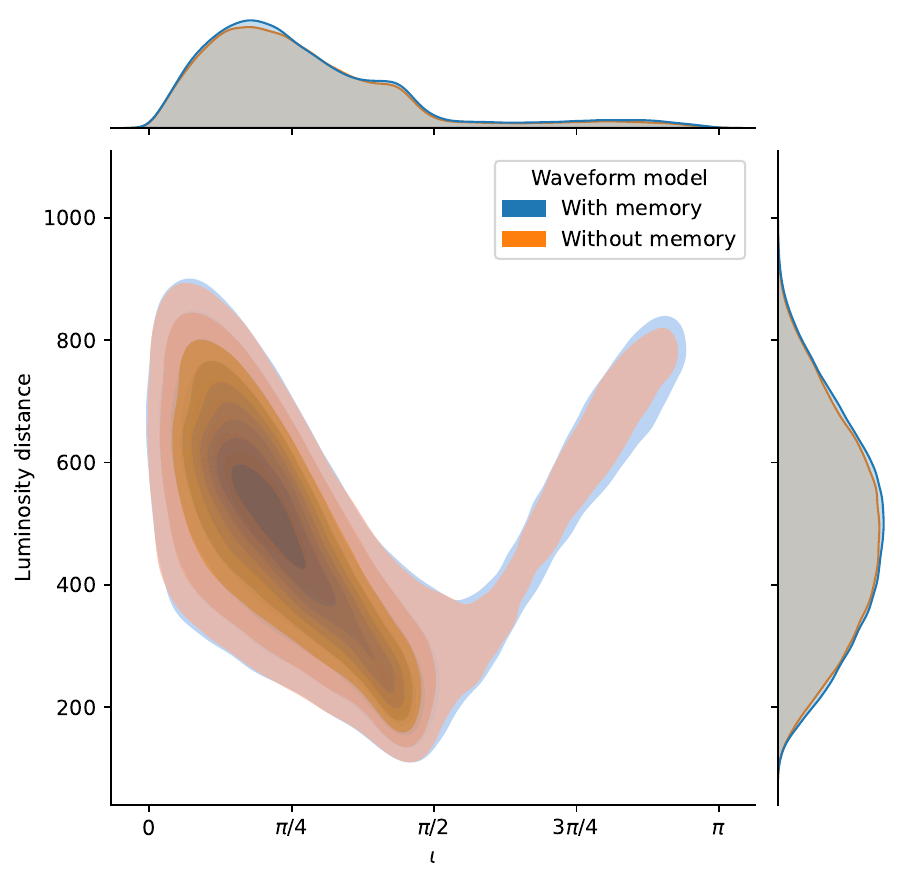}
    }
    \subfloat[\label{fig:snr20_real}]{
        \includegraphics[width=0.35\textwidth]{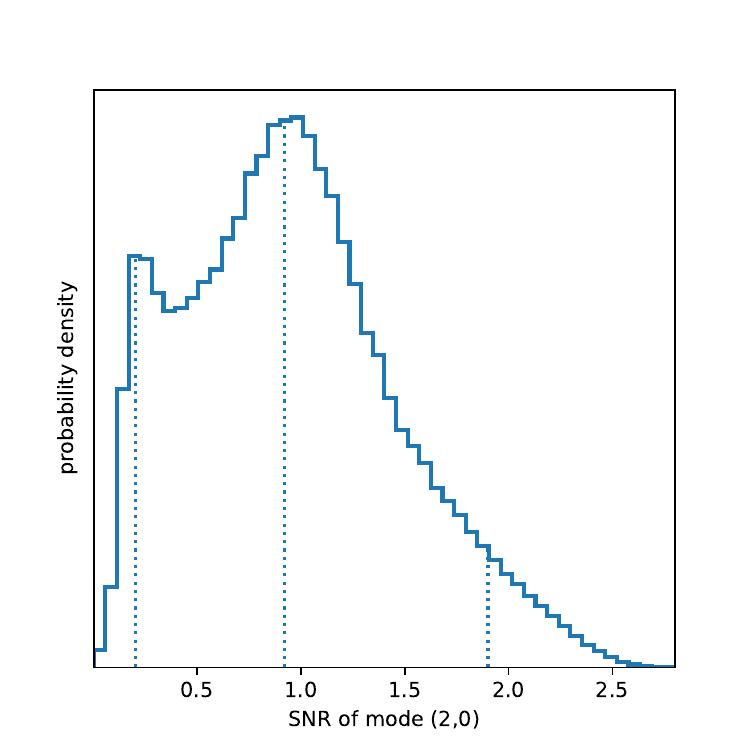}
    }
    \caption{Panel (a) shows the contour plot of inclination and distance posterior of the PE result from GW191204\_171526, with both \texttt{IMRPhenomT\_meme} and  \texttt{IMRPhenomT}. Panel (b) shows the $\rho_{20}^{\perp}$ computed from the posterior samples of PE with \texttt{IMRPhenomT\_meme}.}
\end{figure*}

Although we do not expect non-linear memory to have any impact on detections made during the third observing run of LIGO-Virgo-KAGRA, we would still like to perform an analysis on a real event as a proof of concept and robustness check.
We have selected the event GW191204\_171526 from GWTC-3 \cite{KAGRA:2021vkt}, which has a total mass of 20 $M_\odot$ and an SNR of 17.4. This event was chosen because it has the highest SNR among the low total mass events ($M < 40 M_\odot$), and its mass ratio is around 1.5. 
We used the same prior as the public parameter estimation results \cite{ligo_scientific_collaboration_and_virgo_2023_8177023} except for setting the non-aligned spin to 0. We analyzed the data with both \texttt{IMRPhenomT} and \texttt{IMRPhenomT\_meme}. 

The contour plot of inclination and distance posteriors of the two runs are plotted in Fig.~\ref{fig:joint_real}. As expected, we can see that the inclusion of non-linear memory has no impact on the parameter estimation.

We also computed the $\rho_{20}^{\perp}$ defined in equation \eqref{eq:snr20} in Fig.~\ref{fig:snr20_real} from the posterior samples of the PE with \texttt{IMRPhenomT\_meme}. We see that the 90\% of the credible interval of $\rho_{20}^{\perp}$ is below 1.9, which shows the strength of the non-linear memory in $(2,0)$ mode is not sufficient for this event to affect the PE results.

\section{Conclusions}

In the presented study, we have conducted a detailed investigation into the utility of non-linear memory in (2,0) mode contributed from dominant (2,2) oscillatory mode for improving parameter estimation within the context of gravitational wave astronomy, specifically focusing on its ability to resolve degeneracy in larger parameter spaces using Bayes inference injection methods. Initial assessments with a reduced prior setup were expanded to full-parameter recovery for aligned-spin systems, enabling a comprehensive evaluation of the impact of non-linear memory on enhancing the measurement of luminosity distance and inclination angle --- key parameters for the accurate cosmological or astrophysical studies.
The inclusion of non-linear memory has been shown to offer notable improvements in distance constraints, which is of significant importance for future observations with the upgraded LIGO A$^\sharp$.

Our findings indicate that non-linear memory is particularly effective in breaking the degeneracy between inclination and luminosity distance in high signal-to-noise ratio (SNR $>$ 90) cases for LIGO-O5. The benefit of incorporating non-linear memory is more apparent at closer distances and for non-precessing and symmetric mass systems with higher SNRs. We also note that the (4,4) mode will have similar properties for the symmetric mass system but the inlication angle properties of the (4,4) mode are not as different from the (2,2) mode. The inclusion of (4,4) mode with (2,0) mode will overall improve the inclination angle measurement even for the intermediate values of inclinations as well which is not the case with just (2,0) mode as it improves mostly in the edge-on and face-on systems. The analysis suggests that neglecting non-linear memory effects in high SNR scenarios could lead to systematics in the estimated values of distance and inclination angles.

Further exploration into the implications of LIGO A$^\sharp$’s upgraded sensitivity demonstrates the potential for enhanced parameter estimation accuracy, particularly due to its increased sensitivity at lower frequencies, where non-linear memory effects are more pronounced. The study suggests that with the A$^\sharp$ sensitivity, omitting non-linear memory from the analysis could result in significant systematics in distance measurements for detectable events, even at distances up to about 1 Gpc. 
This is expected to affect almost 30\% of the events in the GWTC-3 catalog based on mean measured distance assuming they were detected by LIGO-A$^\sharp$.
This enhancement is attributed not only to the higher SNR provided by A$^\sharp$ but also to its improved low-frequency sensitivity, crucial for capturing the entirety of non-linear memory effects.

We also applied our methodology to a real gravitational wave event, GW191204\_171526. Despite knowing that its SNR is not high enough to have a significant contribution of non-linear memory in providing better constraints for the accuracy of parameters. We perform the analysis as a proof of concept. The outcome of the analysis is in alignment with the expectations, confirming that non-linear memory's impact is subdued in lower-SNR events without significant changes in the measurement accuracy of other parameters. 

In summary, this research highlights the role of non-linear memory in refining the precision of gravitational wave parameter estimation. As we advance toward the LIGO A$^\sharp$ phase and third-generation detectors, integrating non-linear memory into parameter estimation frameworks will be essential for the accurate localization and characterization of gravitational wave sources. This effort will contribute to the ongoing development of gravitational wave astronomy, enhancing our ability to analyze and interpret the signals from these cosmic phenomena. Additionally, a newly developed phenomenological waveform model for the memory and oscillatory part in the (2,0) mode will facilitate the routine inclusion of memory effects in the future data analysis \cite{Rossello-Sastre:2024zlr}.

\begin{acknowledgments}
We thank Shun Cheung, Paul Lasky, and the referee for their constructive comments.

Y. Xu is supported by the China Scholarship Council.
S. Tiwari is supported by the Swiss National Science Foundation (SNSF) Ambizione Grant Number: PZ00P2-202204.
C. Garci\'a-Quiro\'s is supported by the Swiss National Science Foundation (SNSF) grant Sinergia 213497.
E. Hamilton was supported in part by the Swiss National Science Foundation (SNSF) grant IZCOZ0-189876 and by the UZH Postdoc Grant (Forschungskredit).
Maria Rossell\'o-Sastre is supported by the Spanish Ministry of Universities via an FPU doctoral grant (FPU21/05009).
E. Hamilton, S. Husa and M. Rossell\'o-Sastre were also supported in part by the Universitat de les Illes Balears (UIB); the Spanish Agencia Estatal de Investigaci\'{o}n grants PID2022-138626NB-I00, PID2019-106416GB-I00, RED2022-134204-E, RED2022-134411-T, funded by MCIN/AEI/10.13039/501100011033; the MCIN with funding from the European Union NextGenerationEU/PRTR (PRTR-C17.I1); Comunitat Auton\`{o}ma de les Illes Balears through the Direcci\'{o} General de Recerca, Innovació I Transformaci\'{o} Digital with funds from the Tourist Stay Tax Law (PDR2020/11 - ITS2017-006), the Conselleria d’Economia, Hisenda i Innovaci\'{o} grant numbers SINCO2022/18146 and SINCO2022/6719, co-financed by the European Union and FEDER Operational Program 2021-2027 of the Balearic Islands; the ``ERDF A way of making Europe''.
E. Hamilton is supported in part by the Spanish Ministerio de Ciencia e Innovación (Fondos MRR) - Conselleria de Fons Europeus, Universitat i Cultura with funds from the European Union NextGenerationEU (PRTR-C17.I1) through the project ‘Tecnologías avanzadas para la exploración del universo y sus componentes’ (ref. SINCO2022/6719 ; PI: Alicia Sintes, University of the Balearic Islands, Spain).

The authors are grateful for computational resources provided by Cardiff University supported by STFC grant ST/I006285/1, the LIGO Laboratory supported by National Science Foundation Grants PHY-0757058 and PHY-0823459, and ITP, Lanzhou University supported by National Key R\&D Program of China grant No. 2021YFC2203003.
This paper has document number LIGO-P2400063.

This research has made use of data or software obtained from the Gravitational Wave Open Science Center (gwosc.org), a service of the LIGO Scientific Collaboration, the Virgo Collaboration, and KAGRA. This material is based upon work supported by NSF's LIGO Laboratory which is a major facility fully funded by the National Science Foundation, as well as the Science and Technology Facilities Council (STFC) of the United Kingdom, the Max-Planck-Society (MPS), and the State of Niedersachsen/Germany for support of the construction of Advanced LIGO and construction and operation of the GEO600 detector. Additional support for Advanced LIGO was provided by the Australian Research Council. Virgo is funded, through the European Gravitational Observatory (EGO), by the French Centre National de Recherche Scientifique (CNRS), the Italian Istituto Nazionale di Fisica Nucleare (INFN) and the Dutch Nikhef, with contributions by institutions from Belgium, Germany, Greece, Hungary, Ireland, Japan, Monaco, Poland, Portugal, Spain. KAGRA is supported by Ministry of Education, Culture, Sports, Science and Technology (MEXT), Japan Society for the Promotion of Science (JSPS) in Japan; National Research Foundation (NRF) and Ministry of Science and ICT (MSIT) in Korea; Academia Sinica (AS) and National Science and Technology Council (NSTC) in Taiwan.

\end{acknowledgments}

\appendix

\section{The effect of aligned-spin magnitude}\label{sec:align_spin}

\begin{figure}[h]
    \centering
    \includegraphics[width=0.48\textwidth]{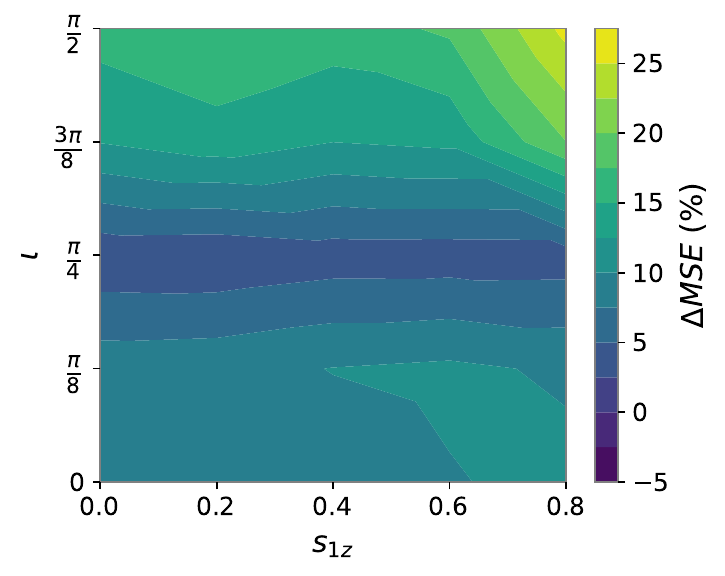}
    \caption{The plot shows the relative fraction improvement  $\Delta \mathrm{MSE}$ for reduce prior, varying the spin magnitude of $s_{1z}$. Twenty-five injections are performed with five inclination angles and five $s_{1z}$ values.}
    \label{fig:spin_mag}
\end{figure}

The amplitude of aligned spin influences the memory amplitude, potentially affecting parameter estimation with memory. In this section, we conducted a series of injections similar to those described in Section \ref{sec:res:reduced_prior}. We injected 25 sets of parameters, comprising 5 values of $s_{1z}$ (evenly spaced from 0 to 0.8) and 5 inclination angles (evenly spaced from 0 to $\pi/2$). The other parameters remain the same as Section \ref{sec:res:reduced_prior} and the aligned spin magnitudes $a_1$ and $a_2$ are varied in the prior. Figure \ref{fig:spin_mag} illustrates the improvement in distance measurement $\Delta \mathrm{MSE}$. The results indicate that for $s_{1z}$ values less than 0.6, the impact on distance measurement is relatively modest with (2,2) mode and its memory contribution in (2,0). However, a substantial improvement in the $\Delta \mathrm{MSE}$ is observed when the spin reaches 0.8. The decrease in $\Delta \mathrm{MSE}$ around $\iota = \pi/4$ is due to the smaller difference compared to the face-on and edge-on cases, with respect to the less informative \texttt{IMRPhenomT} recovery as shown in the upper panel of Fig. \ref{fig:violin_4p}.

\bibliography{ref}

\end{document}